\documentclass[10pt,twocolumn,letterpaper]{article}

\usepackage{cvpr}
\usepackage{times}
\usepackage{epsfig}
\usepackage{graphicx}

\usepackage{amssymb}
\usepackage{amsfonts}
\usepackage{amsmath}
\usepackage{url}
\usepackage{mwe} 
\usepackage{graphicx}
\usepackage{makecell}
\usepackage{multirow,multicol}

\usepackage{url}
\usepackage{xcolor}
\usepackage{hyperref}
\usepackage{pifont}
\definecolor{newcolor}{rgb}{.8,.349,.1}

\cvprfinalcopy 

\def\cvprPaperID{****} 

\begin{document}

\title{A Deep Learning-based Quality Assessment and Segmentation System with a Large-scale Benchmark Dataset for Optical Coherence Tomographic Angiography Image}
%
\author{Yufei Wang\textsuperscript{1,2}, Yiqing Shen\textsuperscript{3}, Meng Yuan\textsuperscript{1}, Jing Xu\textsuperscript{4}, Bin Yang\textsuperscript{5}, Chi Liu\textsuperscript{1,6}, \\Wenjia Cai\textsuperscript{1,*}, Weijing Cheng\textsuperscript{1}, Wei Wang\textsuperscript{1,*}\\
\thanks{Correspondance to W. Cheng and W. Wang. Y. Wang: yufei8828@gmail.com, Y. Shen: shenyq@sjtu.edu.cn, M. Yuan: dr.yuanm@qq.com, J. Xu: xu-j20@mails.tsinghua.edu.cn, B. Yang: yangbin8188@qq.com, C. Liu: chi.liu@student.uts.edu.au, W. Cai: cherylwenjia@outlook.com, W. Cheng: 18120798001@163.com, W. Wang: zoc\_wangwei@yahoo.com.}
\\
\textsuperscript{1} State Key Laboratory of Ophthalmology, \\Zhongshan Ophthalmic Center, Sun Yat-sen University, Guangzhou 510060, China\\
\textsuperscript{2} Key Laboratory of Bio-Resource and Eco-Environment of Ministry of Education, \\College of Life Sciences, State Key Laboratory of Hydraulics \\and Mountain River Engineering, Sichuan University, Chengdu, Sichuan, 610064, China\\
\textsuperscript{3} School of Mathematical Sciences, \\Shanghai Jiao Tong University, Shanghai, 200240, China\\
\textsuperscript{4} Intelligent Computing Laboratory, International Graduate School, \\Tsinghua University, Shenzhen, Guangdong, 518055, China\\
\textsuperscript{5} Department of Ophthalmology, The Third People's \\Hospital of Zigong City, Zigong 643020, China\\
\textsuperscript{6} School of Computer Science, University \\Technology of Sydney, Ultimo NSW 2007, Australia
}
%
%
%
%
%
%
\maketitle
\begin{abstract}
Optical Coherence Tomography Angiography (OCTA) is a non-invasive and non-contacting imaging technique providing visualization of microvasculature of retina and optic nerve head in human eyes \textit{in vivo}. The adequate image quality of OCTA is the prerequisite for the subsequent quantification of retinal microvasculature. Traditionally, the image quality score based on signal strength is used for discriminating low quality. However, it is insufficient for identifying artefacts such as motion and off-centration, which rely specialized knowledge and need  tedious and time-consuming manual identification. One of the most primary issues in OCTA analysis is to sort out the foveal avascular zone (FAZ) region in the retina, which highly correlates with any visual acuity disease. However, the variations in OCTA visual quality affect the performance of deep learning in any downstream marginally. Moreover, filtering the low-quality OCTA images out is both labor-intensive and time-consuming. To address these issues, we develop an automated computer-aided OCTA image processing system using deep neural networks as the classifier and segmentor to help ophthalmologists in clinical diagnosis and research. This system can be an assistive tool as it can process OCTA images of different formats to assess the quality and segment the FAZ area. The source code is freely available at \url{https://github.com/shanzha09/COIPS.git}.

Another major contribution is the large-scale OCTA dataset, namely OCTA-25K-IQA-SEG we publicize for performance evaluation. It is comprised of four subsets, namely sOCTA-3$\times$3-10k, sOCTA-6$\times$6-14k, sOCTA-3$\times$3-1.1k-seg, and dOCTA-6$\times$6-1.1k-seg, which contains a total number of 25,665 images. The large-scale OCTA dataset is available at \url{https://doi.org/10.5281/zenodo.5111975}, \url{https://doi.org/10.5281/zenodo.5111972}. 
\end{abstract}
%
%
\section{Introduction}
The extensive clinical application of Optical Coherence Tomography (OCT) paves the way to the visualization of retinal microvasculature at different layers including superficial, deep, outer layer plexus, and choriocapillaris \textit{in vivo}, rapidly and non-invasively. That is to say, OCT contributes significantly to the clinical diagnosis of retinal diseases, glaucoma, evaluation of cardiovascular, and brain health. OCT angiography (OCTA), a representative technical extension of OCT and has superior microvasculature visualization capability to that of fluorescein angiography, can generate high-resolution retinal vascular morphologic structure images by measuring the blood flow contrast in various retinal layers qualitatively and quantitatively \cite{spaide2015retinal}. The OCTA image of retinal vasculature has been widely utilized in disease staging and preclinical diagnosis, such as diabetic retinopathy \cite{rosen2019earliest}. Moreover, as a novel medical imaging technique, it plays an ever-increasing role in the optical research domain\cite{lauermann2018optical}. Yet, OCTA suffers from the strong image quality variation cased by many factors e.g., patient cooperation, media opacity, optical aberrations, machine-related and etc. Those artifacts of OCTA images such as defocus, shadow, motion artifact, segmentation error, loss of signal, and projection, will cause interpretation difficulty for clinicians \cite{say2017image}. Previous research showed that 31.6\% of conventional tomographic images were ungradable \cite{manjunath2015wide}. 53.5\% OCTA images acquired in multicenter clinical trials had severe artifacts associated with the reliability of quantitative outputs \cite{holmen2019assessment}. Traditionally, image quality score based on signal strength was used to include or exclude OCTA scans for further quantitative analysis. However, the commercially recommended quality index achieved only 37\% to 41\% sensitivity for reliable images \cite{holmen2019assessment}. In addition, artifacts i.e., motion, signal loss, blurriness, and off-centration could not be assessed by signal strength, however manually identification highly relies on well-trained technicians and specialized knowledge. It is a big challenge for manual assessment of OCTA artifacts in busy clinics due to the lack of human resource and insufficient time. 

Another big challenge is the identification of the Foveal Avascular Zone (FAZ) in OCTA images. FAZ is a specialized region of the retina approximating the region of highest cone photoreceptor density and oxygen consumption \cite{jonas1992count,yu2005intraretinal}. The quantitative vascular information provided by the OCTA technique is not only the extremely suited for characterizing the relationships between the morphometric properties of the FAZ and visual acuity (VA) but essential for diagnosing and follow-up the retinal vascular diseases, such as diabetic retinopathy (DR) and retinal vein occlusion (RVO). Previous bio-morphometric analysis indicated that the area of FAZ correlated with VA in DR, RVO, and age-related macular degeneration \cite{balaratnasingam2016visual, salles2016optical, falavarjani2017optical, samara2017quantification, takase2015enlargement}. 
Hence, carrying out the biostatistical analysis of the relationship between FAZ area and optical diseases based on large-scale FAZ area data is on the demand. Yet, the data collection of the FAZ area still remains at the bottleneck stage: limited manual measurement. It has been reported that some OCTA systems have been embedded with automated algorithms for quantifying FAZ metrics. However, the reliability of the systems varies considerably \cite{ishii2019automated, lin2020reliability, magrath2017variability}. On the whole, a creative and reliable technic is on the demand for the clinical diagnosis, to help physicians in unskilled but time-consuming work. 

Deep learning (DL) is a promising technologies, which has received notable attention in medical imaging \cite{pi2020automated, ranjan2017hyperface, yang2017learning,shen2021sampling,shen2021representative}. Currently, DL has been used for assessment of image quality in various medical imaging fields i.e., MRI, CT, ultrasound imaging, fundus photography, and OCT. For example, in the optical image field, some reliable results in OCTA image quality assessment, FAZ segmentation have been reported in previous works. Yet, the generalization ability of DL models depends marginally on the data scale. To be more specific, in OCTA image quality assessment (IQA) tasks, small datasets lead to restricted generalization ability \cite{alam2019supervised, lauermann2019automated}. As for FAZ segmentation, the model was trained on a few high-quality FAZ images so that it cannot be used to quantify the FAZ metrics of most FAZ images \cite{giarratano2020automated, guo2021can,mirshahi2021foveal}. Due to uneven OCTA image quality, automated OCTA image quality assessment and FAZ segmentation cannot be deployed to clinical diagnosis and research widely. 

Considering the ophthalmologists’ suffering of the exhausting manual work, we developed an automated computer-aided OCTA image processing system (COIPS) and addressed the stated challenges through the following methods. (1) We constructed a large-scale OCTA dataset considering four subsets annotated by professional ophthalmologists. This large dataset contributed to improving the generalization of the models. (2) We trained five DL-based image classification models to assess OCTA image quality automatically. Transfer learning was utilized to accelerate the training phase. (3) We trained one U-shape fully convolutional networks to segment the FAZ and quantify the FAZ metrics. (4) We combined the image format conversion, quality assessment, FAZ segmentation, FAZ metrics quantification, and reports generation into a pipeline. This pipeline is what we called COIPS.

The major contributions are summarized as follows.
\begin{enumerate}
 \item Four large-scale OCTA image datasets annotated by professional ophthalmologists was constructed. sOCTA-3$\times$3-10k: 10,480 3 $\times$ 3 $mm^{2}$ superficial vascular layer OCTA (sOCTA) images divided into three classes; sOCTA-6$\times$6-14k: 14,042 6 $\times$ 6 $mm^{2}$ sOCTA images divided into three classes; sOCTA-3$\times$3-1.1k-seg: 1,101 3 $\times$ 3 $mm^{2}$ sOCTA images with pixel-level FAZ annotation; dOCTA-6$\times$6-1.1k-seg: 1,143 6 $\times$ 6 $mm^{2}$ deep vascular layer OCTA (dOCTA) images with  pixel-level FAZ annotation. These OCTA images were labeled and annotated by professional ophthalmologists. All of the images were derived from the healthy subjects or patients with diverse ophthalmic diseases i.e., DR, diabetic macular edema, cataract, age-related macular degeneration and central serous chorioretinopathy in community-based screening in Guangzhou. To our knowledge, these datasets are an order of magnitude lager than any previous public dataset in OCTA-related study.
    \item An automatic COIPS based on DL was developed. According to COIPS configuration file, this system is able to transform OCTA image format, assess octa image quality, segment FAZ, quantify FAZ metrics, and generate the reports of the results automatically, which contributes to reducing the workload of ophthalmologists and saving their time.
    \item Five promising quality assessment models and one FAZ segmentation models were trained on the constructed dataset. Two external testing datasets were introduced to assess the models. Our best 3 $\times$ 3 $mm^{2}$ sOCTA IQA model obtained a weighted accuracy rate of 0.91. Our best 6 $\times$ 6 mm sOCTA IQA model obtained a weighted accuracy rate of 0.90. Our best 3 $\times$ 3 $mm^{2}$ sOCTA images FAZ segmentation model obtained a dice efficient of 0.95. Our best 6 $\times$ 6 $mm^{2}$ dOCTA images FAZ segmentation model obtained a dice efficient of 0.89. Detailed experimental analysis on the performance of these models was done.
    \item The quality assessment and FAZ segmentation results indicate that COIPS has extracted the characteristics of OCTA images successfully. The quality assessment and FAZ segmentation achieved results of COIPS comparable to those of two experienced professional ophthalmologists.
\end{enumerate}
\section{The OCTA-25K-IQA-SEG dataset}
Data-driven methods like deep learning approaches have shown their succeeding success in medical images analysis. Yet, the robustness and generalization ability of deep neural networks, either for classifier or segmenter, highly depends on the training scale. In OCTA analysis tasks, the dataset scale is very restricted to the annotation workload. We list datasets used in literary works in Table \ref{tab1}, describing the tasks they involved in, the available label format, public availability and etc. As it could be observed, only one small-scale OCTA dataset is made publicly available at present. Others including either the superficial vascular layer OCTA (sOCTA) or deep vascular layer OCTA (dOCTA) datasets remains property. To address this issue, one major contribution is our first attempt at constructing and publicizing a large-scale OCTA benchmark dataset, comprising of four subsets, namely sOCTA-3$\times$3-10k, sOCTA-6$\times$6-14k, sOCTA-3$\times$3-1.1k-seg, and dOCTA-6$\times$6-1.1k-seg depending on the image size and format. Moreover, for model evaluation, our datasets provide both the image-level quality assessment annotation and the pixel-level FAZ region mask. All data used in this research have received appropriate ethic approval from Zhongshan ophthalmic center and Sichuan University. In the rest of this section, we elaborate on the construction details, including the curation of all images (c.f. Sec \ref{sec2_1}), our annotation strategies (c.f. Sec \ref{sec2_2}) and the pre-processing procedure (c.f. Sec \ref{sec2_3}). 
\begin{table*}[!t]
\caption{\label{tab1}Details of the OCTA datasets used in previous works. We write ‘QA’ short for ‘quality assessment’ a image level classification task; and FAZS for ‘FAZ segmentation’ a pixel-level segmentation task; 3S short for ‘3 × 3 mm$^{2}$ sOCTA’; 6S short for ‘6 × 6 mm$^{2}$ sOCTA’; 3D represents ‘3 × 3 mm$^{2}$ dOCTA’; 6D writes for ‘6 × 6 mm$^{2}$ dOCTA’. Only one set is made publicly available, denoted by ‘\checkmark’. We make the first attempt to publish a large-scale dataset with labels for both quality assessment and FAZ segmentation.}
\centering
\resizebox{1\linewidth}{!}{
\begin{tabular}{c|c|c|c|c|c}
\hline
Dataset & Dataset scale & \makecell[c]{Field of view \\and image type} & Major task & 
Label category & \makecell[c]{Public \\availability}\\
\hline
\cite{salles2016optical} & 21 & 3S & Correlation between the FAZ and visual acuity & - & $\times$ \\
\cite{guo2018mednet} & 180 & 6S & Avascular area detection & Pixel-level & $\times$ \\
\cite{xu2019automated} & 123 & 6S & QA, FAZS & - & $\times$ \\
\cite{alam2019supervised} & 115 & 6S & Classification of retinopathies & Image-level & $\times$ \\
\cite{lauermann2019automated} & 200 & 3S & QA & Image-level & $\times$ \\
\cite{guo2019development} & 438 & 6S & \makecell[c]{ Nonperfusion area distinguishing\\ artifacts reduction, Pixel-level}&- & $\times$ \\
\cite{giarratano2020automated} & 55 & 3S & Vessel segmentation & Pixel-level &  \checkmark \\
\cite{lin2020improved} & 136 & 3S & FAZS & Pixel-level & $\times$ \\
\cite{gao2019novel} & 180 & 3S & stripe noise removal & - & $\times$ \\
\cite{holmen2020prevalence} & 406 & 3S, 6S & Artifacts analysis & - & $\times$ \\
\cite{guo2021can} & 45 & 3D & FAZS & Pixel-level & $\times$ \\
\cite{mirshahi2021foveal} & 126 & 3S & FAZS & Pixel-level & $\times$ \\
\hline
\textbf{sOCTA-3$\times$3-10k (ours)} & 10480 & 3S & QA & Image-level &  \checkmark \\
\textbf{sOCTA-6$\times$6-14k (ours)} & 14042 & 6S & QA & Image-leve &  \checkmark \\
\textbf{sOCTA-3$\times$3-1.1k-seg (ours)} & 1101 & 3S & FAZS & Pixel-level &  \checkmark \\
\textbf{dOCTA-6$\times$6-1.1k-seg(ours)} & 1143 & 3D & FAZS & Pixel-level &  \checkmark \\
\hline
\end{tabular}
}
\end{table*}

\begin{table*}[!t]
\caption{\label{tab2}Detailed descriptions for each image quality labelled category for the assessment task}
\centering
\resizebox{1\linewidth}{!}{
\begin{tabular}{c|c|c|c}
\hline
Image Quality Category & Image Quality & Macular Clarity& Retinal Vascular Clarity\\
\hline
Ungradable & Insufficient &  Macula may be off-centered or blurring & Blurring \\
Gradable & Moderate blurring or stripe noise & Clear & Moderate blurring \\
outstanding & Fully visible without blurring or with slight stripe noise & Clear & Clear or slight blurring \\
\hline
\end{tabular}
}
\end{table*}

\subsection{OCTA images sources} \label{sec2_1}
The OCTA images used in this study were all collected from Zhongshan ophthalmic center, Sun Tat-Sen university. Specifically, a total number of 25,665 OCTA images were curated from large patient cohorts with general ophthalmic diseases i.e., DR, diabetic macular edema glaucoma, cataract, age-related macular degeneration, and central serous chorioretinopathy. All OCTA scans were digitalized with a swept-source OCTA (SS-OCTA) device (Triton DRI-OCT; Topcon Inc, Tokyo, Japan) under a 1050nm-wavelength-tunable laser at the speed of 100,000 A-scans/s. Images were scaled to 320 $\times$ 320 pixels to obtain high-resolution and visualization en-face images of retinal vascular.

Depending on the image type and the field of view, we distribute whole dataset into four subsets based for study purpose. 1) sOCTA-3$\times$3-10k contains a total number of 10,480 3 $\times$ 3 $mm^{2}$ superficial vascular layer OCTA (sOCTA) images. 2) sOCTA-6$\times$6-14k consists of 14,042 6 $\times$ 6 $mm^{2}$ 6 superficial vascular layer OCTA (sOCTA) images. 3) sOCTA-3$\times$3-1.1k-seg is made up of 1,101 3 $\times$ 3 $mm^{2}$ superficial vascular layer OCTA (sOCTA) images. 4) dOCTA-6$\times$6-1.1k-seg contains 1,143 3 $\times$ 3 $mm^{2}$ deep vascular layer OCTA (dOCTA) images. We did not adopt image augmentation methods in the construction of the dataset in this research to remain the original characteristic of the images.

Images in the external testing data set were curated from West China Hospital, Sichuan University for the performance evaluation.

\subsection{Annotation strategy} \label{sec2_2}
Our datasets provide 2 kinds of annotation: 1) image-level classification label for quality of OCTA images, and 2) pixel-level mask for FAZ region segmentation. 

\begin{figure}[!t]
\centering
\includegraphics[width=1.0\linewidth]{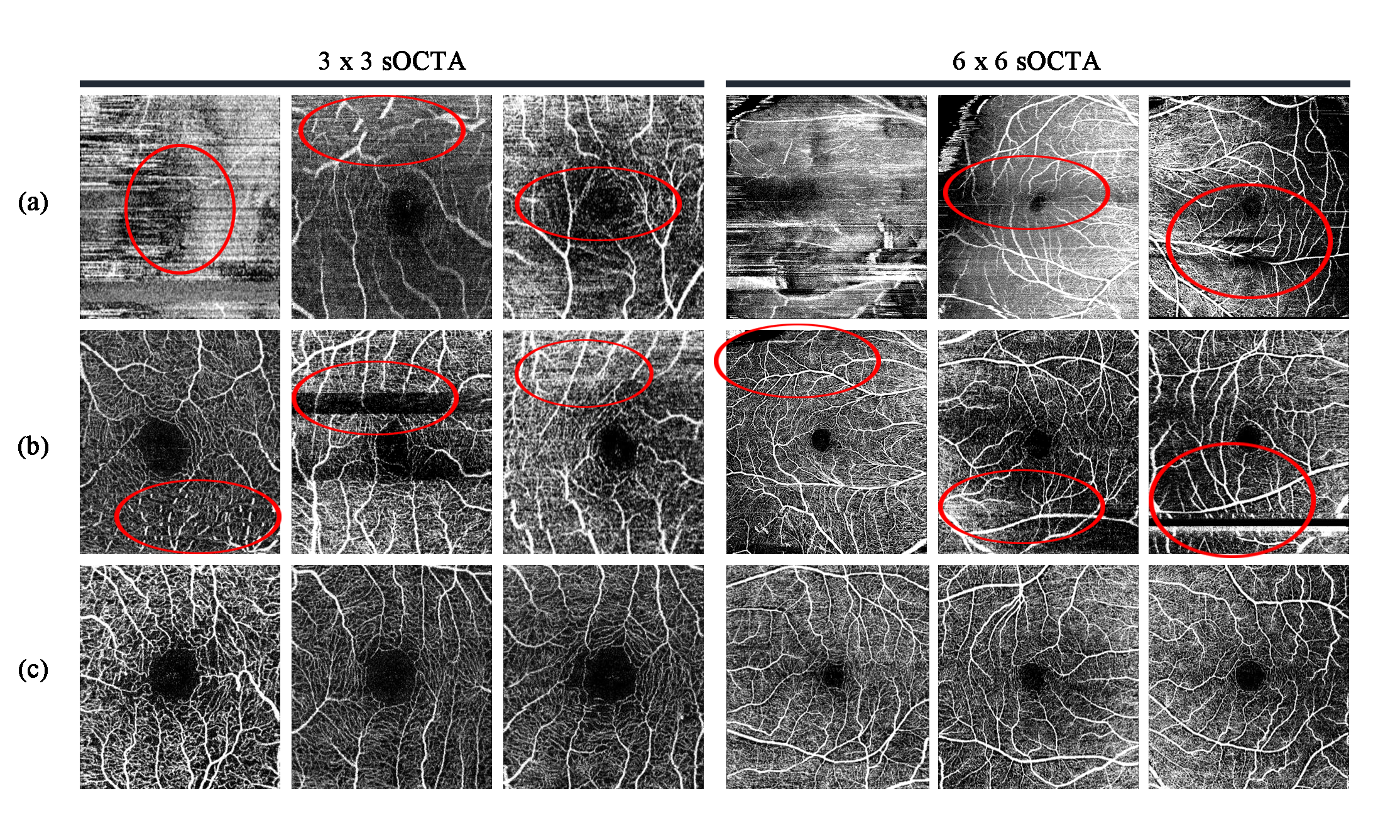} 
\caption{Illustrative OCTA examples from each quality category. (a) Ungradable; (b) Gradable; (c) outstanding. The red circle highlights the blurring region.}
\label{fig1}
\end{figure}

\begin{figure}[!t]
\centering
\includegraphics[width=1.0\linewidth]{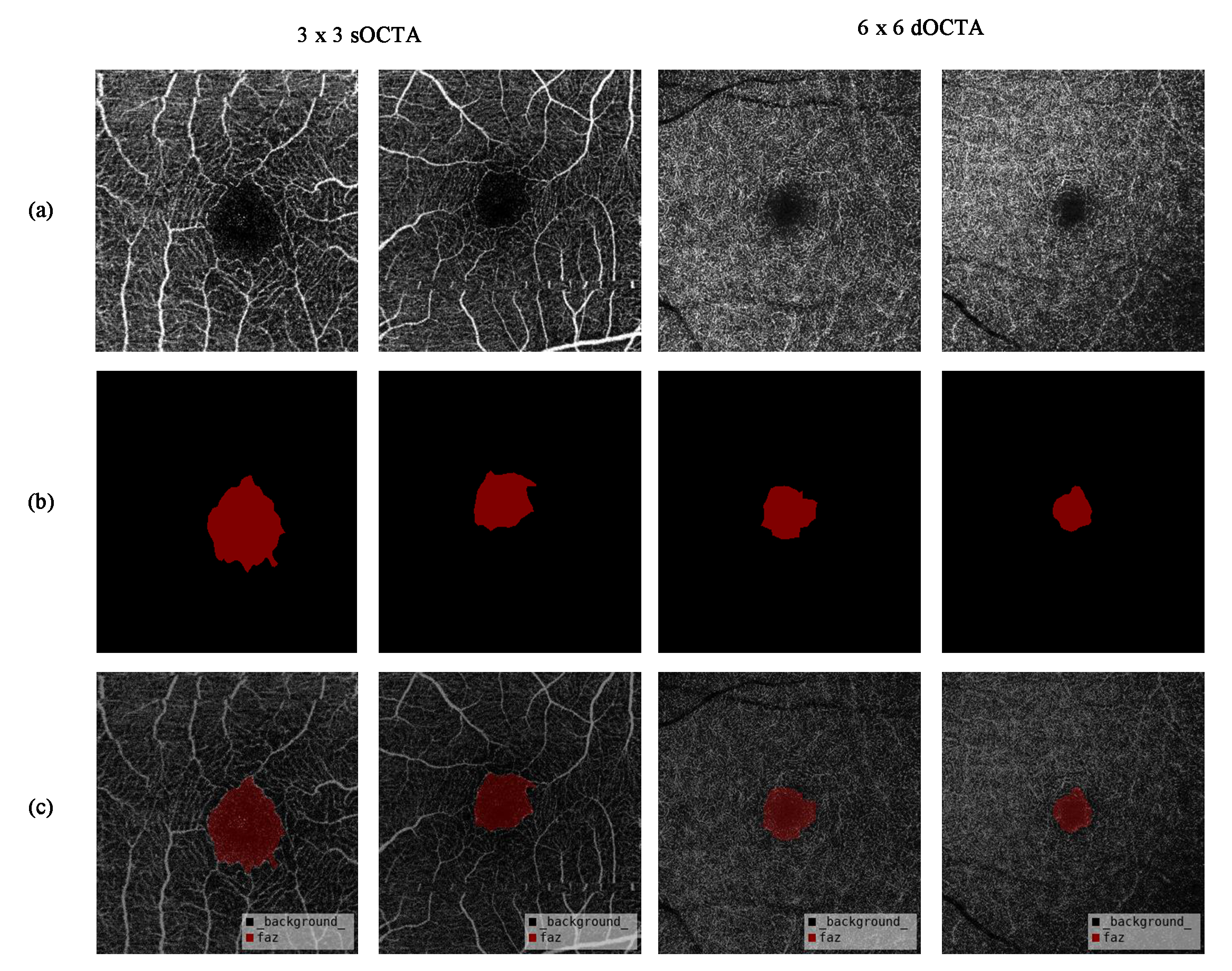} 
\caption{Examples of the OCTA images and the associated FAZ mask. (a) Original OCTA images; (b) Pixel-level ground truth for FAZ; (c) Incorporating annotations to original image.}
\label{fig2}
\end{figure}

The quality assessment is a classification task, where each OCTA image was labeled into one category including i) ungradable, ii) gradable, or iii) outstanding, according to the image quality, macula clarity, and retinal vascular clarity, where the detailed description is in Table \ref{tab2}, and some representative examples illustrated in Fig. \ref{fig1}. All images were labeled by three experienced ophthalmologists independently. The ambigous images were submitted to another senior ophthalmologist to make the final judgement. In annotating pixel-level mask, Labelme (\url{https://github.com/wkentaro/labelme.git}), a graphical image annotation tool, was employed to contour the FAZ in OCTA image. 1,101 3 $\times$ 3 $mm^{2}$ sOCTA images chosen from gradable and outstanding OCTA images randomly in the subset sOCTA-3$\times$3-10k, and 1,143 3 $\times$ 3 $mm^{2}$ dOCTA images were primiparity annotated by three experienced ophthalmologists independently. Illustrative examples for the raw images and the annotated mask can be found in Fig.\ref{fig2}. 

In order to reduce the deviation caused by subjective judgment, 10 percent of the image-level labeled and pixel-level annotated images were selected and checked by another senior experienced ophthalmologist.

\subsection{Data preparation and separation scheme} \label{sec2_3}
Conclusively, a total number of 10,480 3 $\times$ 3 $mm^{2}$ sOCTA images, 14,042 6 $\times$ 6 $mm^{2}$ sOCTA images were labeled at image-level and 1,101 3 $\times$ 3 $mm^{2}$ sOCTA images, 1,143 3 $\times$ 3 $mm^{2}$ dOCTA images were further annotated with pixel-level mask. The labeled 3 $\times$ 3 $mm^{2}$, 6 $\times$ 6 $mm^{2}$ sOCTA images were randomly partitioned into training set, validation set, internal testing set, with hold-out method respectively. Detail descriptions of each set are shown in Table \ref{tab3} and Table \ref{tab4}. For each patient, dOCTA images and sOCTA images are taken simultaneously so that the id of both type images has same prefix. The quality assessment result of sOCTA can be applied to dOCTA.

\begin{table}[!t]
\caption{\label{tab3}The scale of training, validation and testing set for quality assessment task}
\centering
\resizebox{1\linewidth}{!}{
\begin{tabular}{c|c|c}
\hline
\multirow{2}{*}{Sets} & \multicolumn{2}{c}{Quality assessment}\\
\cline{2-3}
&sOCTA-3$\times$3-10k & sOCTA-6$\times$6-14k \\
\hline
Training & 6915 & 9292 \\
Testing & 2965 & 4150 \\
Internal testing & 300 & 300 \\
External testing & 300 & 300 \\
\hline
Total & 10480 & 14042 \\
\hline
\end{tabular}
}
\end{table}

\begin{table}[!t]
\caption{\label{tab4}The dataset scale of training and testing set in FAZ region segmentation}
\centering
\resizebox{1\linewidth}{!}{
\begin{tabular}{c|c|c}
\hline
\multirow{2}{*}{Sets} & \multicolumn{2}{c}{FAZ segmentation}\\
\cline{2-3}
& sOCTA-3$\times$3-1.1k-seg & dOCTA-6$\times$6-1.1k-seg \\
\hline
Training & 708 & 800 \\
Testing & 304 & 343 \\
\hline
Total & 1101 & 1143 \\
\hline
\end{tabular}
}
\end{table}
\subsection{Data available}
We publish our dataset at \url{https://doi.org/10.5281/zenodo.5111975}, \url{https://doi.org/10.5281/zenodo.5111972}.

\section{Methods}
Quality assessment for OCTA images is a primary stage for many downstream analyses including FAZ region location. FAZ region analysis is very essential for real clinical diagnose. To address these issues, a novel automatic OCTA images analysis system is proposed for quality assessment and FAZ segmentation, namely the computer-aided OCTA image processing system (COIPS). In the section, we illustrate the overall framework of our system, and elaborate on each functional component, including a pre-processing stage, a quality assessment block, a FAZ segmentation block, FAZ metrics quantification, and the final aggregation block.
\subsection{Overall framework of the proposed COIPS}
COIPS mainly targets to assist ophthalmologists in grading the raw OCTA images for clinical use. The major motivations are two-fold i.e., 1) quantifying FAZ metrics and 2) generating the prediction based on FAZ region measurement. 

\begin{figure*}[!t]
\centering
\includegraphics[width=1.0\linewidth]{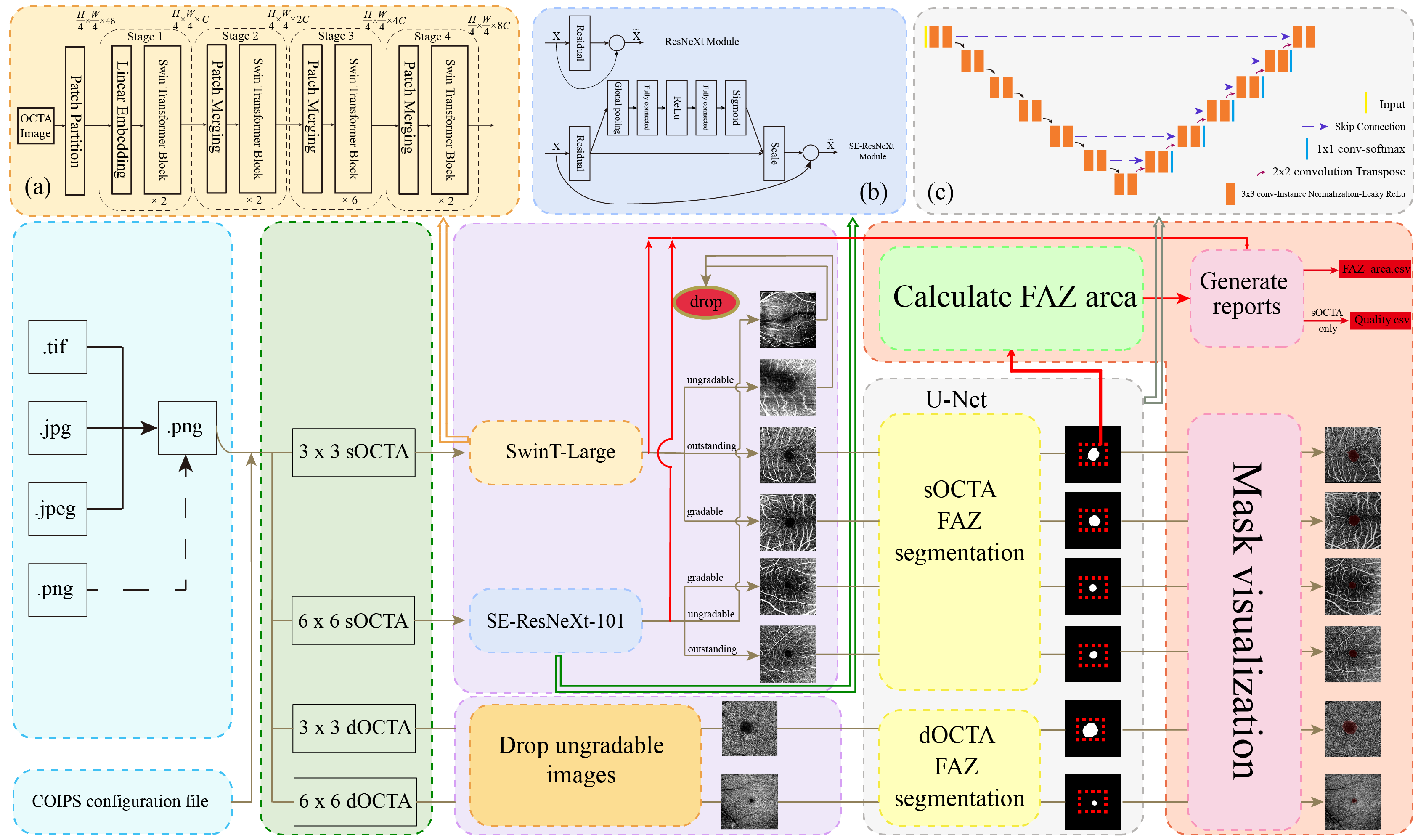}
\caption{Overview of our COIPS. This system can assign an appropriate pipeline based on the input configuration file; (a) The architecture of Swin Transformer (SwinT); (b) Schema of SE-ResNeXt-101 modules; (c) Network architectures generated for FAZ segmentation. OCTA images are first assessed for their visual quality. Based on the  classification result of sOCTA images, the ungradable dOCTA images are discarded, while other images are further segmented for FAZ location. In the next stage, the diagnostic suggestion is derived from an aggregation model based on the segmentation result.}
\label{fig3}
\end{figure*}

To obtain a unified format for image processing, we first covert raw OCTA images to PorTable Network Graphics (PNG) images with OpenCV library (version 4.5.1.48). Afterwards, all OCTA images in png format train a quality assessment block for three-group classification i.e. ungradable, gradable, outstanding. Based on the classification results, images predicted with gradable and outstanding OCTA images will be segmented by a FAZ segmentation U-net. The ungradable images are discarded because these images do not help the disease diagnosis but only harm the computational efficiency. Lastly, the FAZ areas are computed from the segmentation masks and aggregated with a statistic model. Large-scale data of FAZ areas can be obtained through COIPS and be used to characterize the relationship between FAZ areas and optical diseases. The overall framework of the proposed COIPS is shown in Fig. \ref{fig3}.

We adopt image augmentation to obtain a more diverse training pattern, including random horizontal flipping and random rotation. Z-Score normalization of each image is employed to accelerate the training speed and model convergence (Patro and Sahu, 2015), as formulated by:

\begin{equation}
M{_{ij}^C{^\prime} } = \frac{{\left( {M_{ij}^C - \bar X} \right)}}{{{X_\sigma }}} \label{eq1}
\end{equation}
where $M{_{ij}^C{^\prime}}$ is the pixel-level result of Z-Score normalization, $M_{ij}^C$ represents the pixel value of the C-th channel of the original image, $\bar X$ is the mean pixel value of input image $M_{ij}^C$, ${X_\sigma }$ writes for the standard deviation of $M_{ij}^C$.

\subsection{Quality assessment block}
In recent years, medical images classification tasks have benefited greatly from domain adaption including transfer learning, since the restricted quantity of data in the medical domain. For example, to detect and classify the malignant cells in breast cytology images, Khan et al. proposed a CNN-based model which was pre-trained on ImageNet. Then, the model was trained on 6,000 breast cancer histopathology images and achieved an accuracy of 97.525\% on 2,000 breast cancer histopathology images \cite{khan2019novel}. Swati et al. employed transfer learning and they fine-tuned VGG19 to classify the brain tumor for MRI images. Their methods achieved an average accuracy of 94.82\% under five-fold cross-validation with block-wise fine-tuning \cite{swati2019brain}. With the use of fine-tuning on pre-trained models, researchers can accelerate the convergence speed as well as achieve higher generalization ability \cite{zhuang2020comprehensive}. Quality assessment of OCTA images can help ophthalmologists to eliminate the unhelpful artifacts OCTA images. Hence, we also adopt the pre-trained networks as the backbone for three-class classification tasks in quality assessment. 

\begin{figure}[!t]
\centering
\includegraphics[width=1.0\linewidth]{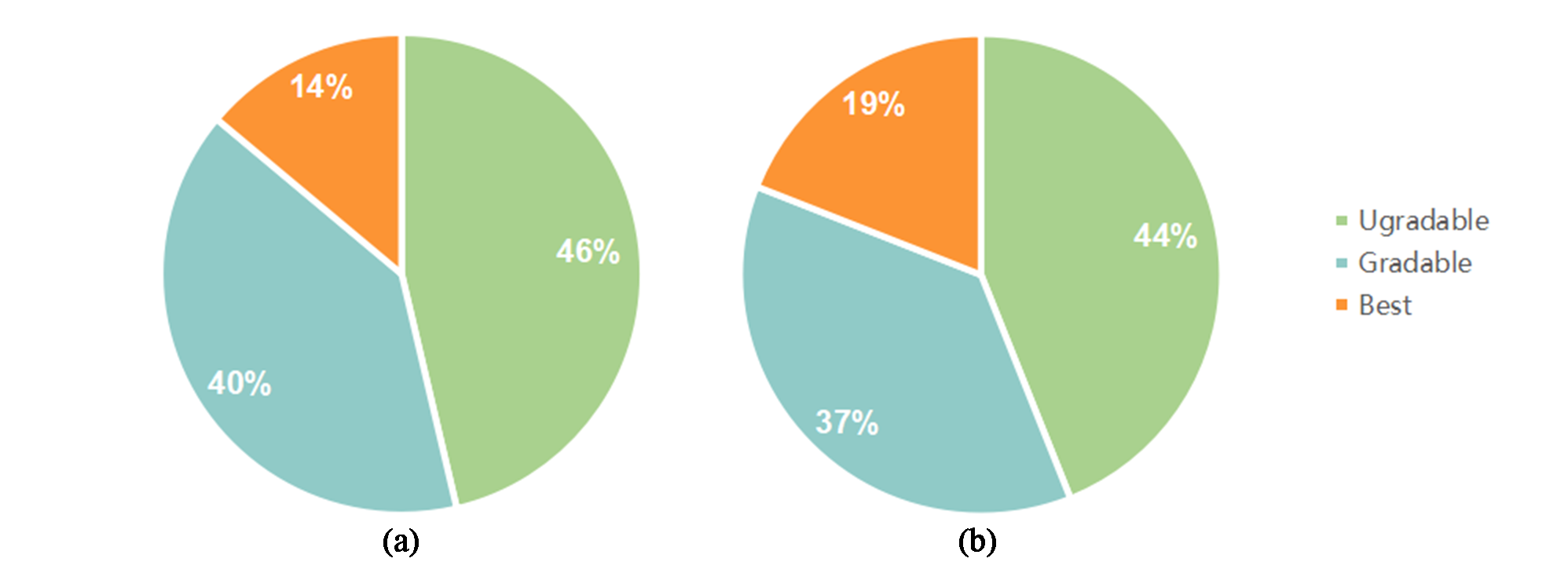}
\caption{Category proportion in two datasets. (a) sOCTA-3$\times$3-10k; (b) sOCTA-6$\times$6-14k.}
\label{fig4}
\end{figure}

The proposed COIPS is CNN-architecture-independent. In this study, five state-of-the-art pre-trained models were adopted as a case study for their outstanding performance on natural image classification and compact architecture. Specifically, we use ResNet-101 \cite{he2016deep}, Inception-V3 \cite{szegedy2016rethinking}, Efficientnet-B7 \cite{tan2019efficientnet}, SE-ResNeXt-101 \cite{hu2018squeeze, xie2017aggregated}, and Swin-Transformer \cite{liu2021swin}  as the backbone feature extractor for assessing the OCTA quality. We fine-tuned the last fully connected layer to fit our dataset, while the parameters in the top layers are frozen. It was a common case that the numbers of data instances from each category were not balanced (shown in Fig. \ref{fig4}), hence a modified cross-entropy loss function  \cite{de2005tutorial} is employed to train our network. It assigns weights to the different categories to focus the target network on the categories with fewer samples i.e.,

\begin{equation}
weigh{t_i} = \frac{{\sum\limits_{i = 1}^N {{X_i}} }}{{N{X_i}}} \label{eq2}
\end{equation}

\begin{equation}
{Y_i} = \left\{ \begin{array}{l}
0,{x_i} \ne {y_i}\\
1,{x_i} = {y_i}
\end{array} \right. \label{eq3}
\end{equation}

\begin{equation}
loss\left( {{x_i},{y_i}} \right) =  - weigh{t_i}\sum {{Y_i}\log \left( {{P_i}} \right)} \label{eq4}
\end{equation}

\begin{equation}
loss = \frac{{\sum\limits_{i = 1}^N {loss\left( {{x_i},{y_i}} \right)} }}{{\sum\limits_{i = 1}^N {weigh{t_i}} }} \label{eq5}
\end{equation}
where $X_i$ is the total number of class $i$, $N$ is the number of category, $x_i$ is the predicted label of class $i$, $y_i$ is the ground truth of class $i$, $P_i$ is the probability of $y_i$.

All the images were resized to 224 $\times$ 224 as the input to ResNet-101, Inception-V3, Efficientnet-B7, SE-ResNeXt-101, 384 $\times$ 384 for Swin-Transformer, and 600 $\times$ 600 for Effcienet-B7. The images were horizontally flipped with a probability of 0.5 and rotated with the angle varied between -15 degrees to 15 degrees randomly for augment.

Prior to the FAZ segmentation in the next section, it is essential to exclude the ungradable images in the quality assessment block. Since these images do not contribute to disease diagnosis and the macular center of them are blurry or even none, but only bring extra computational cost.
\subsection{FAZ segmentation and visualization}
U-Net, an encoder-decoder network architecture, is a primary DL choice for increasing segmentation and quantification tasks and has been the basis of many DL models in biomedical image analysis \cite{ZHOU2021102041mia,falk2019u,ronneberger2015u}. Falk et al. developed an ImageJ plugin based on U-net for cell counting and achieved detection and morphometry with the average intersection over union (IoU) of 85\% \cite{falk2019u}. Guo et al. proposed a customized encoder-decoder network for deep FAZ segmentation with a boundary alignment module and boundary supervision modules and obtained an average Dice of 0.88 \cite{guo2021can}. Isensee et al. reported results on 53 segmentation tasks on 23 public datasets used in international biomedical segmentation competitions and generated 39 state-of-the-art ( sota) segmentation models. These tasks cover a large proportion of the dataset variability in the medical domain \cite{isensee2021nnu}. They also launched a framework, namely nnU-Net. This framework can configure itself automatically with configuration derived by distilling expert knowledge learned from the 53 medical segmentation tasks, which achieved a higher generalization ability than that of models configured on a single dataset. Previous theoretical and empirical results suggest that sophisticated architectural variations are not necessary to obtain a sota model in medical images processing \cite{he2021hf, saood2021covid, zhou2019unet++}.

Accurate FAZ segmentation for OCTA images is the prerequisite for precise quantification of FAZ areas \cite{guo2021can}. Due to various artifacts, conventional medical image analysis techniques such as the denoising filters were not effective enough to the unclear boundary of FAZ, while manual measurement of the FAZ area is unskilled but time-consuming work. In this study, we constructed a U-Net alike architecture based on nnU-Net framework (\url{https://github.com/MIC-DKFZ/nnUNet.git}), as shown in Fig. \ref{fig3} (c). We fine-tuned the architecture to fit our OCTA datasets in segmenting the FAZ region. The network configurations are shown in Table \ref{tab5}. Dice loss defined in Eq. \eqref{eq7} originally targets to maximize the $F_{1}$ Score. The dice loss function is well suited to address the class imbalance, but it will cause unstable training. To achieve better training stability and higher segmentation accuracy, a cross-entropy loss function defined in Eq. \eqref{eq8} was incorporated with dice loss to train the model. 
\begin{equation}
\begin{aligned}
    DiceCoefficient &= \frac{{2\left| {SR \cap GT} \right| + 1}}{{\left| {SR} \right| + \left| {GT} \right| + 1}} \\
   & = \frac{{2TP + 1}}{{2TP + FN + FP + 1}} \label{eq6}
    \end{aligned}
\end{equation}

\begin{equation}
    Los{s_{dice}} = 1 - DiceCoefficient = F_{1} score
    \label{eq7}
\end{equation}
where $SR$ is the segmentation result, $GT$ is the ground truth.  $TP$ denotes true positives (correctly predicted FAZ pixels), $FP$ denotes false positives (incorrectly predicted FAZ pixels), and $FN$ denotes false negatives (incorrectly rejected FAZ pixels), similarly hereinafter.

\begin{equation}
    Los{s_{ce}} =  - \sum {{R_i}\log \left( {{S_i}} \right)} \label{eq8}
\end{equation}
where ${R_i}$ donates the label of pixel $i$, ${S_i}$ donates the prediction of pixel $i$.

To be more specific, the total loss function is formulated as the linear combination of CE-loss and Dice-loss:
\begin{equation}
    Los{s_{Total}} = \lambda \times Los{s_{CE} + (1 - \lambda) \times Los{s_{dice}}} \label{eq9}
\end{equation}
where ${weigh{t_{ce}}}$ and $weigh{t_{dice}}$ are the weight to balance two parts of loss $Los{s_{CE}}$ and $Los{s_{dice}}$. In this study, $\lambda = 0.5$.

To obtain an operating system independent, we re-constructed the nnU-Net framework to run inference on Linux, Windows, and macOS, as provided in the GitHub link. The inferenced masks were visualized with SimpleITK library (version 2.0.2) and Pillow library (version 8.1.2).

\subsection{FAZ metric quantification}
In the this stage, the reports of quality assessment and FAZ metrics were generated for diagnosis. The area of FAZ is calculated as:
\begin{equation}
    {S_{FAZ}} = \frac{{{N_{mask}}{a^2}}}{{{z^2}}} \label{eq10}
\end{equation}
where $N_{mask}$ is the pixel number of predicted FAZ mask, $a$ is the true length at retina that the OCTA image shows, $z$ is the horizontal pixel value of FAZ mask image.

\section{Experiment and results}
In this section, we describe our experimental settings and the empirical results.
\subsection{Platform}
We utilized PyTorch (\url{https://pytorch.org}, version 1.8.1) with Python environment (version 3.9.2) to implement all experiments. All of the trials were carried out using an Ubuntu workstation with an i7-7700k CPU and an NVIDIA GEFORCE RTX 3090 GPU, 64 GB of RAM.

\subsection{Experimental setting-up}
\begin{table*}[!htbp]
\caption{\label{tab5}Network configurations for U-Net}
\centering
\begin{tabular}{c|c}
\hline
Configuration & Assigned Parameter \\
\hline
Batch size & 64 \\ 
Patch size & [512, 512] \\
Number of pooling per axis & [7, 7] \\ 
Down-sampling strides & [[2, 2], [2, 2], [2, 2], [2, 2], [2, 2], [2, 2], [2, 2]] \\
Convolution kernel sizes & [[3, 3], [3, 3], [3, 3], [3, 3], [3, 3], [3, 3], [3, 3], [3, 3]] \\
\hline
\end{tabular}
\end{table*}
Quality assessment: All the weights of models used in this study were pretrained on the ImageNet dataset \cite{deng2009imagenet}. All of the training run for a fixed length of 300 epochs. To save computing resources and time, the program was early stopped when the test loss was no longer reduced within 20 epochs. We also warmed up the training with the learning rate of 0.0000005 for 20 epochs when trained Swin-Transformer-Large model. As for optimizer, Adam with an initial learning rate of 0.001 was used for training ResNet101, Inception-V3, SE-ResNeXt101, and Efficinet-B7 \cite{kingma2014adam}. AdamW with an initial learning rate of 0.0005 was used for training Swin-Transformer-Large \cite{loshchilov2018fixing}. The learning rate was reduced during training using “CosineAnnealingLR” schedule with “T\_MAX” of 5 \cite{loshchilov2016sgdr}. 

FAZ segmentation: Stochastic gradient descent (SGD) was chosen as the optimizer with an initial learning rate of 0.01 and Nesterov momentum of 0.99 \cite{Bottou2012}. 5-fold-cross-validation was trained and the best one was used to infer. We employed the “poly” learning rate (ployLR) policy as a learning rate reducing schedule with the exponent of 0.9 \cite{chen2017deeplab}. All of the other configuration parameters were generated by the nnU-Net framework, automatically.
\subsection{Evaluation metrics}
Quality assessment: We adopted a variety of measures commonly used for classification tasks to assess the performance of the quality assessment models, including accuracy (Acc), precision (Pre), area under the receiver operating characteristic curve (AUC), $F_{1}$ score, and confusion matrix. The model with the highest validation accuracy was selected as the backbone classifier for FAZ segmentation evaluation. 

FAZ segmentation: We employ the following metrics: dice coefficient, Jaccard index, precision and recall. Jaccard index, precision and recall are defined as: 
\begin{equation}
    {J_{index}} = \frac{{\left| {GT \cap SR} \right|}}{{\left| {GT \cup SR} \right|}} \label{eq11}
\end{equation}

\begin{equation}
     precission = \frac{{TP}}{{TP + FP}} \label{eq12}
\end{equation}

\begin{equation}
     recall = \frac{{TP}}{{TP + FN}} \label{eq13}
\end{equation}

\begin{figure}[!htbt]
    \centering
    \includegraphics[width=1.0\linewidth]{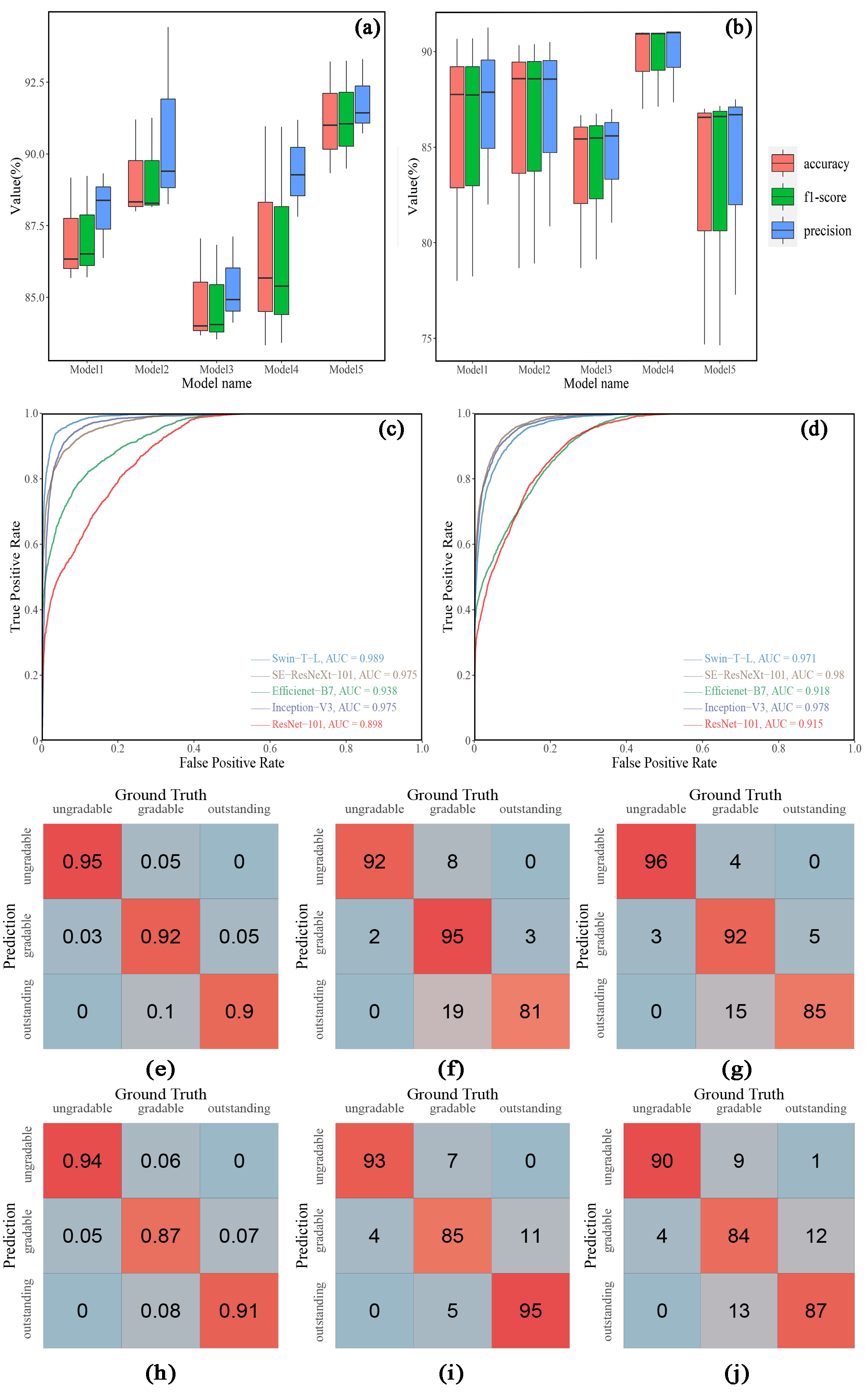}
    \caption{Results visualization of all of the models. (a) – (b) Box-plot of accuracy, precision and $F_{1}$ score of different models for 3 $\times$ 3 $mm^{2}$ sOCTA (left) and 6 $\times$ 6 $mm^{2}$ sOCTA (right) quality assessment; (c) – (d) ROC curve of different models for 3 $\times$ 3 $mm^{2}$ sOCTA (left) and 6 $\times$ 6 $mm^{2}$ sOCTA (right) quality assessment; (e) – (g) Confusion Matrix plot of predictions of testing set (left), internal testing set (middle) and external testing set (right) using Swin-transformer-Large to assess 3 $\times$ 3 $mm^{2}$ sOCTA images; (h) – (j) Confusion Matrix plot of predictions of testing set (left), internal testing set (middle) and external testing set (right) using SE-ResNeXt-101 to assess 6 $\times$ 6 $mm^{2}$ sOCTA images. Where ‘Model1’ is short for ‘Efficienet-B7’, ‘Model2’ is short for ‘Inception-V3’, ‘Model3’ is short for ‘ResNet-101’, ‘Model4’ is short for ‘SE-ResNeXt-101’, ‘Model5’ is short for ‘Swin-Transormer-Large’.}
    \label{fig5}
\end{figure}

\subsection{Results and visualization}
Quality assessment: Table \ref{tab6} demonstrates the classification performance of quality assessment models. More detailed evaluations including Acc, Pre, and $F_{1}$ score on each sub-dataset are shown in Supplementary Note, Table S1 – S10. The box-plot figures in Fig. \ref{fig5}, (a) – (b) suggests that Swin-Transformer-Large outperforms other classification models on 3 $\times$ 3 $mm^{2}$ sOCTA quality assessment; and SE-ResNeXt-101 stands out on 6 $\times$ 6 $mm^{2}$ sOCTA quality assessment in this study. Specifically, Swin-Transformer-Large achieves an accuracy of 0.91, a precision of 0.92, and an $F_{1}$ score of 0.91 on 3 $\times$ 3 $mm^{2}$ sOCTA testing data set, marginally succeeds the other four network architectures. SE-ResNeXt-101 achieve an overall accuracy of 0.90, a precision of 0.90 and a $F_{1}$ score of 0.90 on 6 $\times$ 6 $mm^{2}$ sOCTA testing data set. The Receiver Operating Characteristic (ROC) curve evaluates the sensitivity of the classification models, are illustrated in Fig. \ref{fig5}, (c) – (d). The ROC curve figures of each category are shown in Supplementary Note, Fig. S1 – S10. It is observable that Swin-Transformer-Large and SE-ResNeXt-101 performs better than other models in terms of 3 $\times$ 3 $mm^{2}$ , 6 $\times$ 6 $mm^{2}$ sOCTA quality assessment, respectively. Both of Swin-Transformer-Large and SE-ResNeXt-101 can achieve an Auc of 0.98. The confusion matrix shows a precise number of predictions for each dataset in Fig. \ref{fig5}, (e) – (j). It can be seen that there are no significant differences in prediction performance among the three categories. Therefore, we used Swin-Transformer-Large and SE-ResNeXt-101 to assess the quality of 3 $\times$ 3 $mm^{2}$, 6 $\times$ 6 $mm^{2}$ sOCTA images, respectively for the downstream FAZ segmentation task.

\begin{figure}[!htbt]
    \centering
    \includegraphics[width=1.0\linewidth]{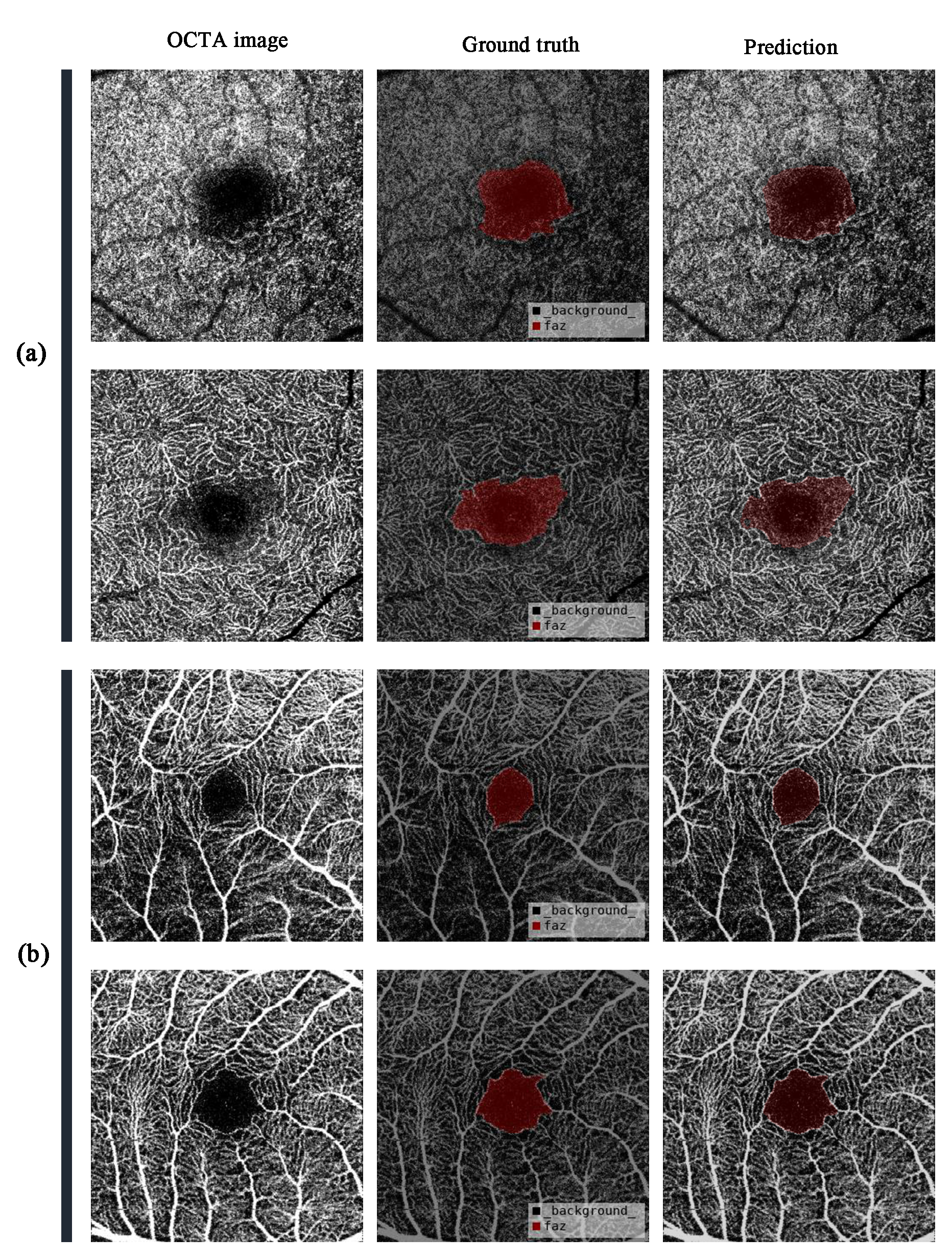}
    \caption{Segmentation results of 3 $\times$ 3 $mm^{2}$ OCTA images against the ground truth. (a) Prediction of dOCTA images; (b) Prediction of sOCTA images.}
    \label{fig6}
\end{figure}

\begin{figure}[!htbt]
    \centering
    \includegraphics[width=1.0\linewidth]{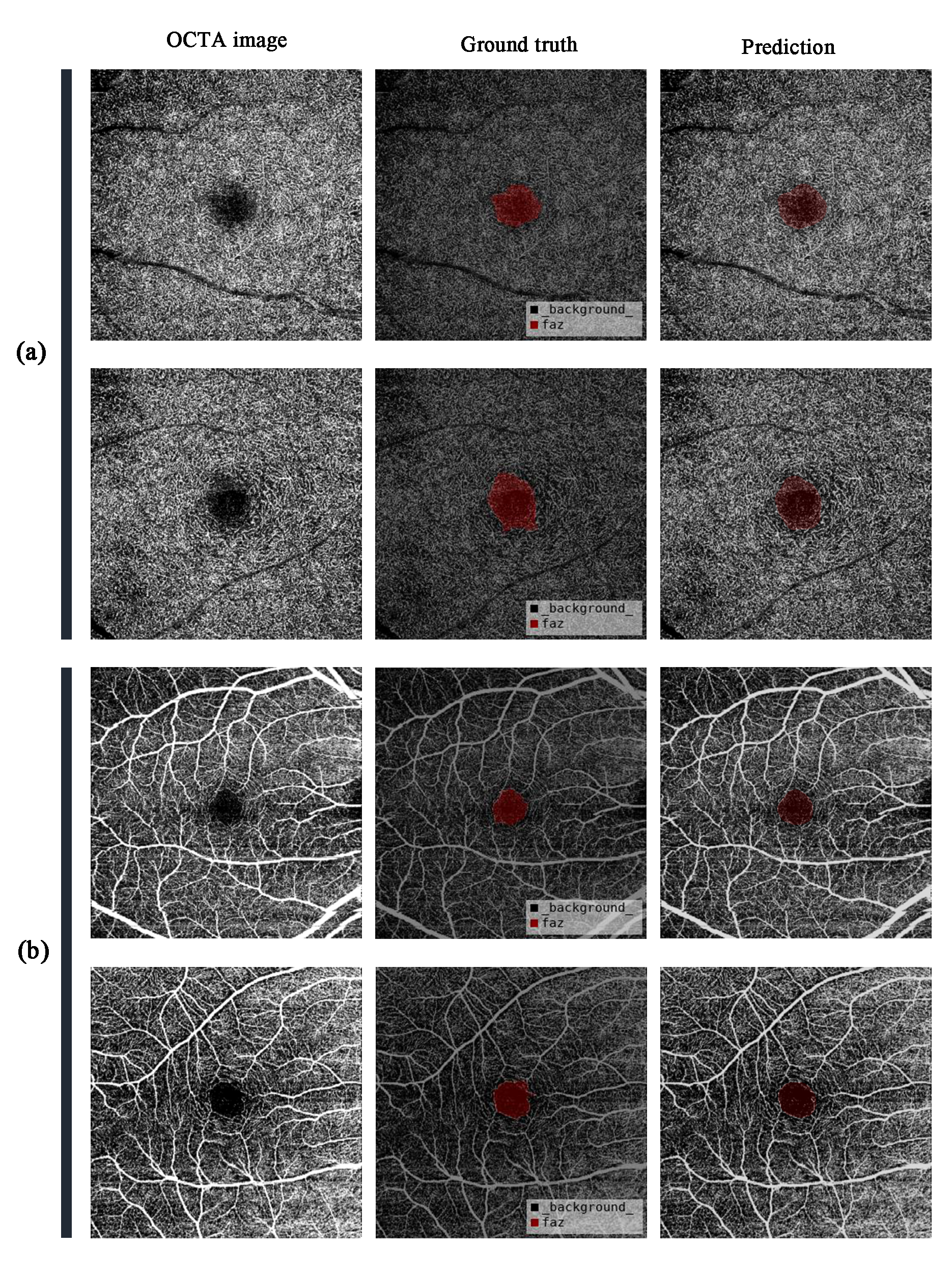}
    \caption{Segmentation results of 6 $\times$ 6 $mm^{2}$ OCTA images against the ground truth. (a) Prediction of dOCTA images; (b) Prediction of sOCTA images.}
    \label{fig7}
\end{figure}

FAZ segmentation: The overall performance of the FAZ segmentation model on two datasets is shown in Table \ref{tab7}. The model trained on sOCTA-3$\times$3-1.1k-seg achieves a dice coefficient of 0.95, a Jaccard index of 0.91, a precision of 0.96, and a recall of 0.95. Trained on dOCTA-6$\times$6-1.1k-seg dataset, the model achieved a dice coefficient of 0.89, a Jaccard index of 0.79, a precision of 0.89, and a recall of 0.89. We visualized the FAZ segmentation mask on two subsets compared with ground truth in Fig. \ref{fig6} and Fig. \ref{fig7}. As shown in the figures, the model can locate the FAZ area accurately with a smooth boundary.

\section{Conclusion and Discussion}
In recent years, the OCTA technique shows promising results in getting the images of the smallest capillaries from the retina without the necessity of a contrast agent \cite{giarratano2020automated}. However, the pathologic information OCTA images reflect highly depends on the quality of the image. To address this issue, we construct an effective image-level quality assessment neural network classifier. Besides, FAZ segmentation of OCTA images shows great significance to clinical diagnosis as well as for academic research propose.

\begin{table*}[!htbt]
\caption{\label{tab6}Overall performance of quality assessment models. The number in the Table is the average of testing set, internal testing set and external testing set, respectively. where 3 $\times$ 3 is short for ‘3 $\times$ 3 mm$^{2}$ sOCTA’; 6$\times$6 is short for ‘6 $\times$ 6 mm$^{2}$ sOCTA’.}
\centering
\begin{tabular}{c|c|c|c|c|c|c|c|c|c|c}
\hline
\multirow{2}{*}{Metrics} &\multicolumn{2}{c|}{ResNet-101} & \multicolumn{2}{c|}{SE-ResNeXt-101} & \multicolumn{2}{c|}{Efficientnet-B7 }&\multicolumn{2}{c|}{ Swin-T-Large }& \multicolumn{2}{c}{Inception-V3 }\\
\cline{2-11}
& 3$\times$3 & 6$\times$6 & 3$\times$3 & 6$\times$6 & 3$\times$3 & 6$\times$6 & 3$\times$3 & 6$\times$6 & 3$\times$3 & 6$\times$6 \\
\hline
Acc & 0.85 & 0.80 & 0.87 & 0.90 & 0.87 & 0.85 & 0.91 & 0.82 & 0.89 & 0.86 \\
Pre & 0.85 & 0.85 & 0.89 & 0.90 & 0.88 & 0.87 & 0.92 & 0.84 & 0.90 & 0.87 \\
$F_{1}$ score & 0.85 & 0.84 & 0.87 & 0.90 & 0.87 & 0.86 & 0.91 & 0.83 & 0.89 & 0.86 \\
Auc & 0.90 & 0.91 & 0.96 & 0.98 & 0.93 & 0.92 & 0.98 & 0.96 & 0.97 & 0.97 \\
\hline
\end{tabular}
\end{table*}

\begin{table}[!htbp]
\caption{\label{tab7}Overall performance of trained FAZ segmentation model on two datasets}
\centering
\resizebox{1\linewidth}{!}{
\begin{tabular}{c|c|c}
\hline
Metrics & sOCTA-3$\times$3-1.1k-seg & dOCTA-6$\times$6-1.1k-seg \\
\hline
Dice coefficient & 0.95 & 0.89 \\ 
Jaccard index & 0.91 & 0.79 \\
Precision & 0.96 & 0.89 \\
Recall & 0.95 & 0.89 \\
\hline
\end{tabular}
}
\end{table}
While the existing solutions for FAZ quantification including built-in programs, conventional images processing techniques, and manual measurement, are labor-intensive and computationally costly. Furthermore, limited by methods of FAZ area measurement, there is little research on the correlation between FAZ area and retinal vascular diseases. Therefore, we propose a novel automated COIPS in this research to address all these issues. This system can serve as assistive tools to ophthalmologists by reducing their many workloads. Our system also shows large generalization ability to be extended to all storage format OCTA images by conversion into unified PNG format for processing, assess and classify the images, segment and quantify FAZ and report the results automatically. Compared with existing research \cite{giarratano2020automated, lauermann2019automated, lin2020improved, mirshahi2021foveal}, our methods achieve outstanding performance, which suggests that this system can provide a convincing result of OCTA quality assessment as well as FAZ segmentation. 

Another major contribution is that we make the first attempt at constructing a public available large-scale dataset containing four sub-datasets, namely sOCTA-3$\times$3-10k, sOCTA-6$\times$6-14k, sOCTA-3$\times$3-1.1k-seg, and dOCTA-6$\times$6-1.1k-seg. To our best knowledge, the dataset is the largest one in this field up to now. We trained each of these five OCTA IQA models: ResNet-101, Inception-V3, Efficienet-B7, SE-ResNeXt-101, and Swin-Transformer-Large on sOCTA-3$\times$3-10k and sOCTA-6$\times$6-14k dataset and evaluated the performance on the testing set, internal testing set, and external testing set, respectively. For 3 $\times$ 3 $mm^{2}$ sOCTA quality assessment, Swin-Transformer-Large did better than other models and achieved an accuracy of 0.91, a precision of 0.92, and an $F_{1}$ score of 0.91. For 6 $\times$ 6 $mm^{2}$ sOCTA quality assessment, SE-ResNeXt-101 did better than other models and achieved an accuracy of 0.90, a precision of 0.90, and an $F_{1}$ score of 0.90. In this study, the number of each category per dataset varies greatly. A modified cross-entropy loss function was employed to train the models. The results show that there are no significant differences in prediction performance among the three categories. In view of the importance of FAZ, we trained a U-Net-like architecture based on nnU-Net framework to segment FAZ and quantify the area. The FAZ segmentation models obtained a dice of 0.95 based on sOCTA-3$\times$3-1.1k-seg dataset and a dice of 0.89 based on dOCTA-6$\times$6-1.1k-seg dataset. The different performance of the model on two datasets may be caused by the different features of the OCTA images. The boundary of 6 $\times$ 6 $mm^{2}$ OCTA images is not as clear as 3 $\times$ 3 $mm^{2}$ OCTA images, which makes FAZ area location more difficult. To sum up, the results indicate that deep learning methods are competent for classification and segmentation jobs upon OCTA images and have a broad scope in ophthalmology. 

The COIPS we developed made quality assessment and FAZ area quantification very efficient, rendering it practicable for clinical applications. In the future, promising improvements could be achieved by the following aspects: (1) evaluate and verifying COIPS on a new dataset in a clinical setting; (2) inference stage performance speed up; (3) fine-tune the feature extractor to get a higher precision of each category.






\bibliography{main}
\bibliographystyle{IEEEbib}
\end{document}